# Failure Mechanism Traceability and Application in Human System Interface of Nuclear Power Plants using RESHA

**Edward Chen[a], Han Bao[b], Tate Shorthill[c], Carl Elks[d], Nam Dinh[e]**
[a]North Carolina State University, 2500 Stinson Dr., Raleigh, NC, 27607, USA, echen2@ncsu.edu
[b]Idaho National Laboratory, P.O. Box 1625, MS 3860, Idaho Falls, ID 83415, USA, han.bao@inl.gov
[c]University of Pittsburgh, 3700 O'Hara Street, Pittsburgh, PA 15261, USA, ths60@pitt.edu
[d]Virginia Commonwealth University, Richmond, VA, USA, crelks@vcu.edu
[e]North Carolina State University, 2500 Stinson Dr., Raleigh, NC, 27607, USA, ntdinh@ncsu.edu

**Abstract:** In recent years, there has been considerable effort to modernize existing and new nuclear power plants with digital instrumentation and control systems (DI&C). However, there has also been considerable concern both by industry and regulatory bodies for the risk and consequence analysis of these systems. Of particular concern are digital common cause failures (CCFs) specifically related to software defects. These "misbehaviors" by the software can occur in both the control and monitoring of a system. While many new methods have been proposed to identify potential software failure modes, such as Systems-theoretic Process Analysis (STPA), Hazard and Consequence Analysis for Digital Systems (HAZCADS), etc., these methods are focused primarily on the control action pathway of a system. In contrast, the information feedback pathway lacks unsafe control actions (UCAs), which are typically related to software basic events; thus, assessment of software basic events in such systems is unclear. In this work, we present the idea of intermediate processors and unsafe information flow (UIF) to help safety analysts trace failure mechanisms in the feedback pathway and how they can be integrated into a fault tree for improved assessment capability. The concepts presented are demonstrated in two comprehensive case studies, a smart sensor integrated platform for unmanned autonomous vehicles and another on a representative advanced human system interface (HSI) for safety critical plant monitoring. The qualitative software basic events are identified, and a fault tree analysis is conducted based on a modified Redundancy-guided Systems-theoretic Hazard Analysis (RESHA) methodology. The case studies demonstrate the use of UIF and intermediate processors in the fault tree to improve traceability of software failures in highly complex digital instrumentation feedback. The improved method can also clarify fault tree construction when multiple component dependencies are present in the system.

## 1. INTRODUCTION

The versatility and capability of software on digital platforms is especially attractive to the nuclear industry, which has routinely relied on analog counterparts. However, single and common cause software failures[*] have become an increasing issue, especially for digital instrumentation and control (DI&C) systems [1]. Systems theoretic process analysis (STPA) [2] is one of the more novel and well-known methods to identify software failure modes. Specifically, it identifies the type of loss scenarios that software control systems can cause and the corresponding unsafe control actions (UCAs). A system must have "authority" over a physical process to enact control actions (e.g., sensors do not have authority over the process they are monitoring). Control actions become unsafe when they occur within particular contexts or conditions resulting in a defined loss [2]. It is important to note that UCAs are failure modes of the software and are focused on the control mechanisms available. These UCAs can also be integrated into a fault tree to assess causality between software basic events and top events. This concept was introduced in Hazard and Consequence Analysis for Digital Systems (HAZCADS) [3], where a hardware fault tree is constructed to include relevant UCAs. The REdundancy-guided Systems-theoretic Hazard Analysis (RESHA) [4] expands on the concepts from HAZCADS to inform on the construction of integrated software fault trees for highly redundant DI&C systems [5]. RESHA guides the structure of the integrated fault tree and the identification of common cause failure (CCF) basic events through explicit consideration of the redundant architecture of the DI&C system's divisions and modules. In STPA, the authors discuss the failure mechanisms behind UCAs, including an inadequate control and process model, unsafe control input, and inadequate feedback and information [2]. In Figure 1, failure mechanisms of both the actuation and feedback pathways can be seen.

---
[*] In this work, 'failure' is used generally as an undesirable outcome by the system, which includes unintended design or 'misbehaviors.'

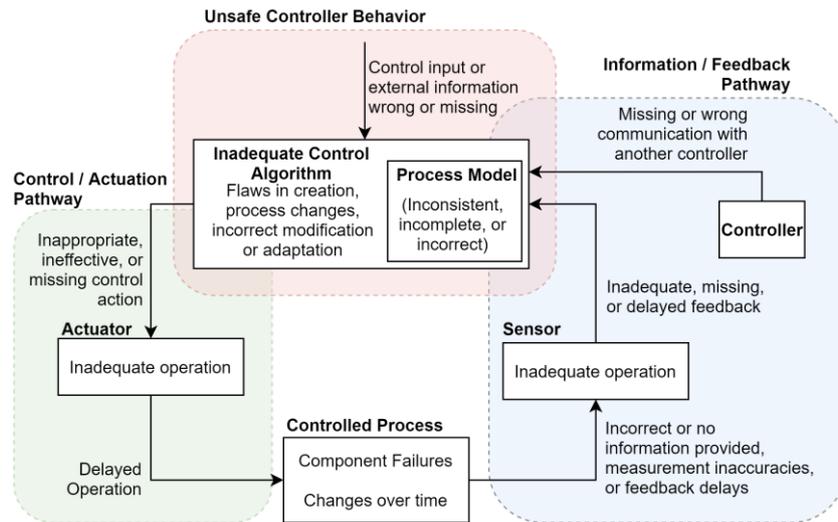

Figure 1. Failure mechanisms in the feedback and actuation pathway for UCAs adaptive from [2].

However, tracing failure mechanisms from UCAs can be convoluted in complex systems. This problem is further exacerbated in control-free systems whose only functionality includes the monitoring of a physical process. For example, in the Advanced Power Reactor 1400 (APR-1400) [6], reactor monitoring includes the Qualified Indication and Alarm System for Safety (QIAS-P), a safety critical system with no capability of controlling the reactor. Within a division of the QIAS-P, there are multiple intermediate feedback systems that parse, validate, and augment the data before forwarding. Expressing traditional UCA failure modes in this control-free system is difficult. Tracing the failure mechanisms is also difficult without defined failure modes. These problems also extend to conventional nuclear control systems, such as the reactor trip system (RTS) and engineered safety features and auxiliary systems (ESFAS) where control is dependent on various sensors and feedback systems. The purpose of this work is to clarify how failure modes and mechanisms of complex information/feedback pathways can be classified and organized to show their impacts to system safety. Ultimately, we want to identify which areas of the DI&C system contribute the most risk to operations, whether it is inadequacies in the controller software or due to external factors, such as failures in dependencies and sensors.

In this work, we propose a method to support the traceability of failure mechanisms in the information/feedback pathway with an integration into fault trees. Specifically, unsafe information flow (UIF) is proposed to capture how monitoring and sensing systems can fail. A detailed description of UIF is provided, why UIFs are needed, and how they can be integrated into a fault tree. We also propose a revised method to structure fault trees to reflect internal and external failure mechanisms. RESHA is partially utilized in this work for the integrated fault tree construction. While this work is intended for safety critical monitoring and control-free systems, the method can also be applied to any system with dependencies on information from digital sources.

This work is part of the Light Water Reactor Sustainability (LWRS) project, "DI&C Risk Assessment". The object of this project is to provide effective quantitative and qualitative measurement tools to gauge the risk and reliability of modernization projects in existing and novel nuclear reactors [7] [8] [9].

## 2. THEORETICAL BACKGROUND

Before presenting the approach, the theory of software failure mechanisms and modes are discussed. Failure modes can be broadly defined as the way which a unit/component fails [10] or went wrong within a system. In HAZCADS [3], UCAs are added to fault trees as software failure modes as actions that can cause defined losses under worst-case scenarios. In contrast, failure mechanisms are ways for the failure mode to occur and describe the transitory states of the system [10], or simply the failure cause. Under the same loss scenarios, failure mechanisms are the cause of UCAs. The failure mechanisms that are the focus of this work are failures in the information/feedback system, known in this work as UIFs. Another common term that arises is root cause. Importantly failure mechanisms are not equivalent to root causes. A root cause describes the exact set of conditions or inadequacies in the system resulting in a failure [11]. There can be multiple root causes for the same failure mechanism as how multiple failure mechanisms can lead to the same failure mode. For example, a failure mode may be that a controller fails to provide an action when needed. The failure mechanisms behind that failure mode could include issues such as a controller failed to receive feedback from dependent devices or sensors. The root cause for this failure mechanism could be that the communication protocol between the two devices was configured incorrectly. In this work, we do not identify root causes behind failure modes. Instead, we categorize failure mechanisms into distinct groups that describe what can happen.

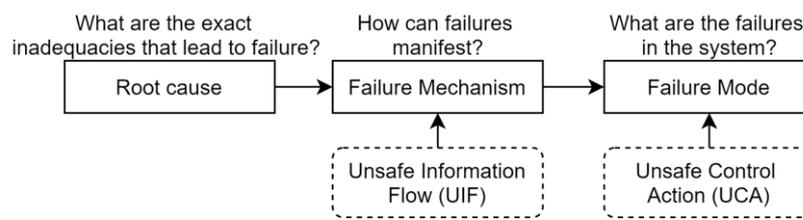

Figure 2. Relationship between root cause, failure mechanism and modes, and UIF and UCAs.

Mechanisms that can trigger UCAs, described [2], can be broadly categorized into two groups. Internal factors such as inadequacies in the control algorithm or process model, and external factors such as inadequate feedback and information input from dependencies. In this first group, failure mechanisms manifest due to two primary deficiencies. The first is an incomplete design process where not all relevant constraints, requirements, or conditions needed for intended operation are considered. The second deficiency is an incomplete engineering/implementation process, where hidden assumptions or human errors in the software coding process results in residual defects. Hardware failures of the software platform are also included in the first group (i.e., integrated circuit burnout). The second group of failure mechanisms manifest in the transference of knowledge between systems. Specifically, these include failures in the data transmission infrastructure, such as input-output ports, wires, sensors, excess noise, malicious spoofing, etc., as well as failures in dependent hardware and software devices. While mechanisms are identified in [2] and [3], the tracing of these mechanisms to specific DI&C components was not considered in the scope of their work. The primary difference between the mechanism groups is the location of the inadequacy. In group one, the failure is from an internal defect, while in group two, the defect is external. This is used later to organize events in our fault tree.

For example, assume a digital pressurizer controller is implemented to control pressure in a pressurizer via spray nozzles. The inputs to the controller include a digital pressure value transmitted from an analog to digital signal converter (ADC). The ADC itself is a digital integrated circuit that receives analog information from pressure sensors. When examining the pressurizer controller, potential UCAs that may lead to over-pressurization of the system could include: (1) The controller does not provide the SPRAY ON control signal to the spray nozzles when high system pressure is detected; or (2) The controller provides SPRAY ON signal to the spray nozzles too late when high system pressurization is detected. These UCAs are failure modes of the system because they describe what went wrong. The failure mechanisms for these UCAs could include internal mechanisms, such as 'spray controller turn ON setpoint too high,' but also external mechanisms, such as software failures in the ADC (i.e., incorrect conversion of pressure analog signal results in seemingly normal values). Note, the mechanism describes a scenario that leads to a failure under a defined loss scenario, but is not a failure by itself.

The UIFs are organized into four categories based on failures in feedback: (UIF-A) failure to provide information when needed; (UIF-B) providing information when not needed; (UIF-C) providing information but is either early, late, not in sync, or out-of-order; and (UIF-D) providing information that is too low or too high, not-a-number (NaN), or infinity (inf). UIF-B can most easily be related to actuation systems that send information sporadically as needed (e.g., alarms, notification, polling systems). UIF-C and UIF-D are related to continuous monitoring systems that transmit state information (e.g., real-time smart sensors). UIF-A can include both types of systems when state information or signals are missing. Chaining of UIF events is also a valid construction as failures in one information system can lead to failures in other information systems. In general, UIFs can only exist if UCA failure modes are possible (e.g., human controllers can also cause UCAs). UIFs must be traceable to the controller that ultimately consumes the information, regardless of the number of layers between the UIF and the UCA. UIFs that cannot be traced to a defined failure mode also cannot result in control failures. For example, a monitoring system with no controller or human to report to also has no influence over the physical process. In the pressurizer example, a lack of a controller would render UIF modeling of the ADC pointless because the physical process state would not change regardless of the proper or improper output of the ADC.

In this work, the ADC is referred to as an intermediate processor. Specifically, they are physical subsystems that do not retrieve raw analog information (e.g., voltage difference from thermocouple) or has authority for control. Instead, intermediate processors only convert, augment, or modify information for other subsystems. While the pressurizer example only had one intermediate processor, in systems with multiple redundant and diverse intermediate processors (e.g., QIAS-P), the tracing of all software failure mechanisms can be convoluted. Intermediate processors are only used to refine the scope of analysis and the output signals of concern. An AD41111 is an example of a specific ADC device that can interpreted as an intermediate processor. Without this constraint, the tracing of information feedback failures can become convoluted—especially for software variables. Realistic complex software systems may have hundreds of arbitrary variables that feed into dependent functions; modeling each would only unnecessarily complicate the assessment.

## 3. METHODOLOGY

As mentioned previously, UIFs are identified based on the physical separation of hardware components for intermediate processors. A top-down systems analysis approach (i.e., STPA) is used to deconstruct the control or monitoring system into relevant subsystems and components. Here, it is important to identify the major hardware components, inputs and outputs of each module, and dependency across components. For general purpose monitoring systems, there may be multiple outputs from the physical system. Each variable is assumed to have equal importance in the analysis and treated as a batch. Any incorrect signal in the batch is assumed to trigger the corresponding UIF (i.e., 1 out-of 5 output signals is missing when needed is considered a UIF-A event).

The UIF types are described and follow similar categories as the UCAs. In table 1, the original UCAs are provided for comparison against UIF formulation. In UCA-A, a control action is missing when needed, causing a hazard. UIF-A is similar where feedback is missing when needed, causing a hazard. This event is applicable to both discrete (i.e., alarms) and continuous (i.e., real-time sensor) monitoring systems. In UCA-B, a control action is provided when not needed, causing a hazard. UCA-B is an example of a spurious actuation. A UIF-B occurs when feedback about a process is provided unexpectedly. For example, an alarm triggers spuriously when the physical state is normal. UIF-B is not applicable for continuous monitoring as feedback in such systems will always be needed. Rather, in continuous monitoring systems, Type C and Type D failures are more applicable. In UIF-C, the feedback timing is incorrect or out-of-sync with the physical process. This can occur when the timestamps reported by the system are not aligned with actual time. With UCA-D, the control action is stopped too soon or applied too long. For information systems, this is equivalent to the feedback value being incorrect. For example, the received sensor value is lower than it should be or if the system provides unintelligible feedback, for instance Not a Number (NaN), infinity (Inf), or nonsensical feedback ('garbage'). Duplicate UIFs across redundant divisions are treated as common cause software failure events. After relevant modules and UIFs have been identified, they can be included in the integrated fault tree.

Table 1. Comparison between Unsafe control actions and information flows.

|  | Type A | Type B | Type C | Type D |
|---|---|---|---|---|
| **Unsafe Control Action (UCA)** | Control action not provided causing hazard. | Control action provided causing hazard. | Control action is early, late, or out-of-order. | Control action is stopped too soon or applied too long. |
| **Unsafe Information Flow (UIF)** | Feedback is missing when needed causing hazard. | Feedback is provided when not needed causing hazard. | Feedback is early, late, out-of-sync, or out-of-order. | Feedback value is too low, too high, NaN, or Inf. |

The construction of integrated fault trees follows a similar methodology, as described in RESHA. For brevity, the major steps in the method are provided. First, define the top event targeted for assessment. Next, identify hardware modules that can lead to this top event and identify dependencies. The construction of the initial fault tree is thus a hardware fault tree consisting of hardware failure basic events. Two empty branches are also assigned—a placeholder software failure branch and a dependency branch. The software failure branch is intended to include only controller software failure modes (UCAs) caused by Group 1 internal failure mechanisms. For intermediate processors, the UCAs under this branch are replaced with UIFs caused by a Group 1 internal failure mechanism. The dependency branch includes components that the current component depends on (i.e., other intermediate processors, sensors, other controllers, etc.). Group 2 external failure mechanisms to the current component also are added under this branch. By sufficient iteration, all software basic events can be represented as Group 1 internal failure mechanisms. In Figure 3, an example of the integrated fault tree is provided.

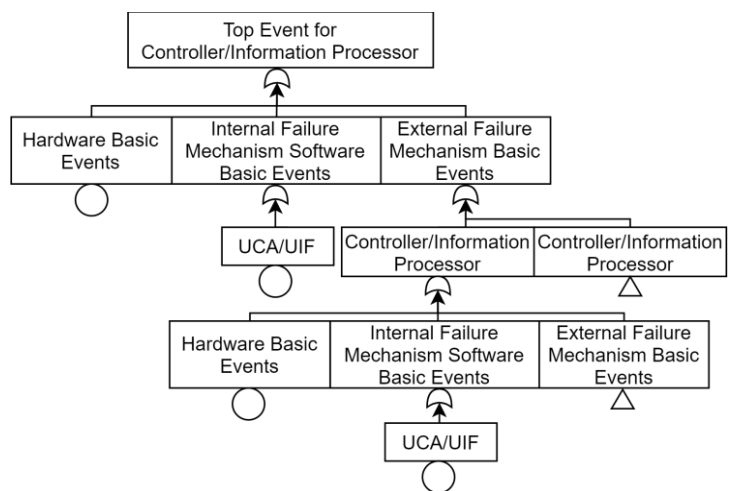

Figure 3. Sample construction of an integrated fault tree with the inclusion of intermediate processors, and UCAs/UIFs through separation of the internal and external failure mechanisms.

One concern that fault tree developers may have is the double counting and organization of basic events. As UIF failure mechanisms lead to UCA failure modes, it would make more logical sense to organize the tree where UIFs lead to UCAs, as observed in the left side of Figure 4. UCA-A and UCA-B are caused by the same dependent device, which, due to an internal defect, can be caused by either UIF-A or UIF-B. By this configuration, it would suggest that UIF-A and UIF-B are CCFs to UCA-A and UCA-B. This is not possible for single controllers as control actions are mutually exclusive (i.e., a controller cannot complete two actions concurrently). Furthermore, the UCAs are no longer basic events, and the organization of the tree must be significantly modified. The separation between the internal and external failure mechanisms also becomes convoluted. Finally, while the two trees in Figure 4 seem to be different, following the idempotent addition law in fault tree construction [11], the two trees are actually equivalent when calculating the top event. An important assumption here is that UIFs will always trigger UCA failure modes under the worst conditions. This assumption is reasonable as controller dependencies that fail will trigger a hazardous condition for the controller, thus initiating a UCA. Thus, it is recommended that the right proposed fault tree configuration is utilized. The immediate benefit is the separation between failures caused by internal and external mechanisms, which offers better clarity when tracing events. It allows analysts to compare the number of internal and external factors that can cause UCAs, thereby focusing on which areas of the system need improvements. Furthermore, dependency branches can be added to existing fault trees without modifying already identified UCA failure modes. This allows for UIFs to be incorporated into many existing methods like RESHA and HAZCADS. Ultimately, the goal is to understand which areas of the DI&C control system introduces more risk to operation (e.g., controller software vs. dependencies) [12].

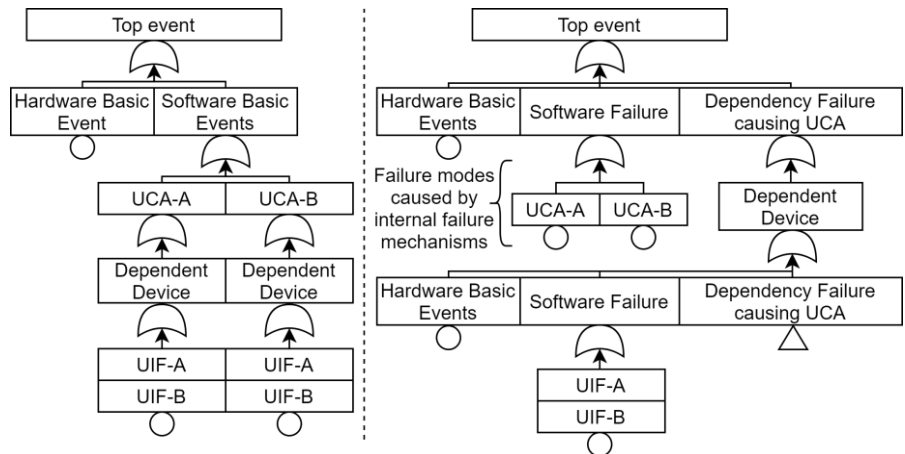

Figure 4. Left side, potential logical organization of fault tree. Right side, proposed configuration of fault tree

## 4. CASE STUDY

A demonstration of the proposed method is conducted on two separate systems. The first system is a smart sensor platform [13] with a lightweight operating system (ChibiOS) [14]. The ChibiOS operating system is an example of an all-purpose multi-use resource management system. The ChibiOS is also compatible with multiple different hardware platforms and does not have a dedicated hardware to separate exact functionality. The ChibiOS software was used to control various functionalities of the smart sensor including filtering and processing of the analog sensor values. The main processor for the Virginia Commonwealth University (VCU) sensor—STM32F4 ARM Cortex-M4—can be used for a variety of functions and can compute numerous intermediate variables. A conventional assessment of the operating system hardware or software would be difficult as tracing all the used arbitrary variables would be highly labor-intensive. The sensor software (including ChibiOS) and hardware was integrated as a physically separate module on the ARIES_2 Advanced Autopilot Platform for autonomous drone control. The main intended outputs from the sensor are pressure and temperature, which are used by the ARIES system to determine true altitude, which is then used to correct to the desired altitude. This case study is used to demonstrate how failure mechanisms in generalized hardware architecture and complex software can be modeled with UIFs. An integrated fault tree is developed based on the overall system description.

The second case study is on a representative human system interface (HSI) modeled after the QIAS-P from the APR-1400. Note that this HSI is listed as a safety critical infrastructure for reactor trip safety. In addition, the HSI presented hereafter is not a true replica and mentioned components are assumed from public documentation. All specifications pertaining to the QIAS-P were derived from the publicly available design information (see reference [6]). While the exact design of the system is proprietary, sufficient detail was provided to approximate the structure of the system. This case study does not constitute a complete safety assessment but demonstrates how various interweaving and dependent modules can be modeled in a highly complex information retrieval system. The QIAS-P was implemented to provide safety critical information and alarm notification on the reactor state through various sensors such as the heated junction thermocouples (HJTCs) and core exit thermocouples (CETs). Within the system, there are two redundant divisions (e.g., A and B), which are assumed architecturally identical, as diversity was not specified. Within each division, there are various value calculators that determine intermediate variables, such as the reactor coolant saturation margin (RCSM). There also exists diverse monitoring systems such as the Qualified Indication and Alarm System for Non-Safety (QIAS-N), the Diverse Instrumentation System (DIS), and information processing system (IPS). All systems have independent arrays of sensors and are physically isolated from each other. In this paper, only one control action is assessed for the entire monitoring system which is the operator manual reactor trip. It is likely the system(s) also influences other control actions but, in this work, only actions relevant to reactor trip safety were considered. No automated trip safety features can be influenced because the monitoring systems were intended to be independent from the existing automatic RTS system and provide an alternative reading of the physical plant state. This case study is used to demonstrate how multiple layers of intermediate processors, within a complex multi-division DI&C architecture, can be modeled following UIF principles.

**4.1 VCU Smart Sensor Integrated Fault Tree Analysis**

The VCU smart sensor's primary function is to provide readings for the adequate control of altitude [13]. The assumed top event for this assessment is assigned as, 'Autonomous drone crashes into ground due to incorrect sensor reading." The corresponding hardware devices relevant to this top event are now identified. In Table 2, the various relevant hardware components and outputs are described. Note the STM32F4, while called a 'microcontroller' in this system, does not control a physical process. For clarification, digital data is represented as a square wave with pre-programmed voltage levels for binary 1s and 0s. Analog data is represented as a floating voltage level between zero and the max output voltage. The first three sensors measure the various pressure and temperature readings required by the system. The MS4525DO outputs digital data via I2C communication protocol to the microcontroller. The MP3H6115A and MP3V5004DP both output analog data but are soldered on the same circuit board. For the STM microcontroller, only the onboard ADC is specified. Other integrated circuits related to the STM32FM407 could have been further elaborated on but the direct impact to the top event was unclear and thus excluded. The selection of relevant hardware components is at the discretion of the safety analyst.

**Table 2. Relevant hardware for assessment for the VCU smart sensor.**

| Component ID | Function | Output Type |
|---|---|---|
| MS4525DO | Absolute or differential pressure and temperature sensor. | Digital I2C data |
| MP3H6115A | Static pressure sensor (no software). | Analog voltage level |
| MP3V5004DP | Dynamic pressure sensor (no software). | Analog voltage level |
| ADC (STM32) | Analog to digital converter for microcontroller input channels. | Digital data |
| STM32FM407 | General purpose micro, calculates pressure, temperature, and Kalman-filtered pressure. | Digital UART data |

First, the hardware fault tree is constructed from Table 2 and shown visually in Figure 5. Empty branches associated with the software failures and the dependency failures are included for each identified digital component. The digital components in the VCU smart sensor include the MS4525DO, ADC, and STM32FM407. Under the top-level software failure branch, is a UCA of concern directly related to the top event. Failure mechanisms under the software failure branch include internal mechanisms specific to the controller. Primarily, failures in the controller algorithm due to inadequate program requirements or failures in the process model. Under the dependency branch, the VCU smart sensor platform and all other hardware dependencies of the sensor are listed. For instance, the smart sensor depends on the MS4525DO, MP3H6115A, and MP3V5004DP sensors for state awareness. It also depends on the ADC embedded on the STM32F4M407 to convert analog signals to readable digital signals. The potential UIFs that can lead to the top event are if any of the readings are reported as higher than true (UIF-D). This is possible in the ADC converter, the MS4525DO sensor head, and in the UART communication from the STM32FM407. The qualitative information that can be drawn from the fault tree (Figure 5) are thus the various UIF feedback failure mechanisms in the sensory system that can lead to the top event. An example of a UIF occurring could be the default output state of the MS4525DO showing absolute pressure and instead of differential pressure. If the STM32FM407 assumes that the MS4525DO is reporting differential pressure, the system may falsely report that the change in pressure is always positive. This would ultimately inform the altitude controller incorrectly when conducting adjustments. Note that the operating system, ChibiOS, that implements the software is not included in the fault tree. This is because an operating system cannot be uniquely identified by a single component but is a composition of various integrated circuits instead. However, the exclusion of the operating system does not detract from the overall usefulness of the assessment. All intended outputs of the system (e.g., temperature, pressure, differential pressure) can be modeled without explicitly modeling the operating system. If the intention of the VCU smart sensor is to offer reliability like in nuclear reactor systems, the immediate conclusion drawn from the qualitative assessment is the need for additional device redundancy. Specifically, for improved resiliency against single failures, an additional STM32FM407 board can be implemented in parallel. In addition, a changing the software to treat the three onboard sensors as redundant and diverse backups to each other would improve resiliency against CCF. For example, the MP3H6115A and MP3V5004DP both sense pressure in different ways and may be used as diverse protection against CCF software failures of the MS4525DO.

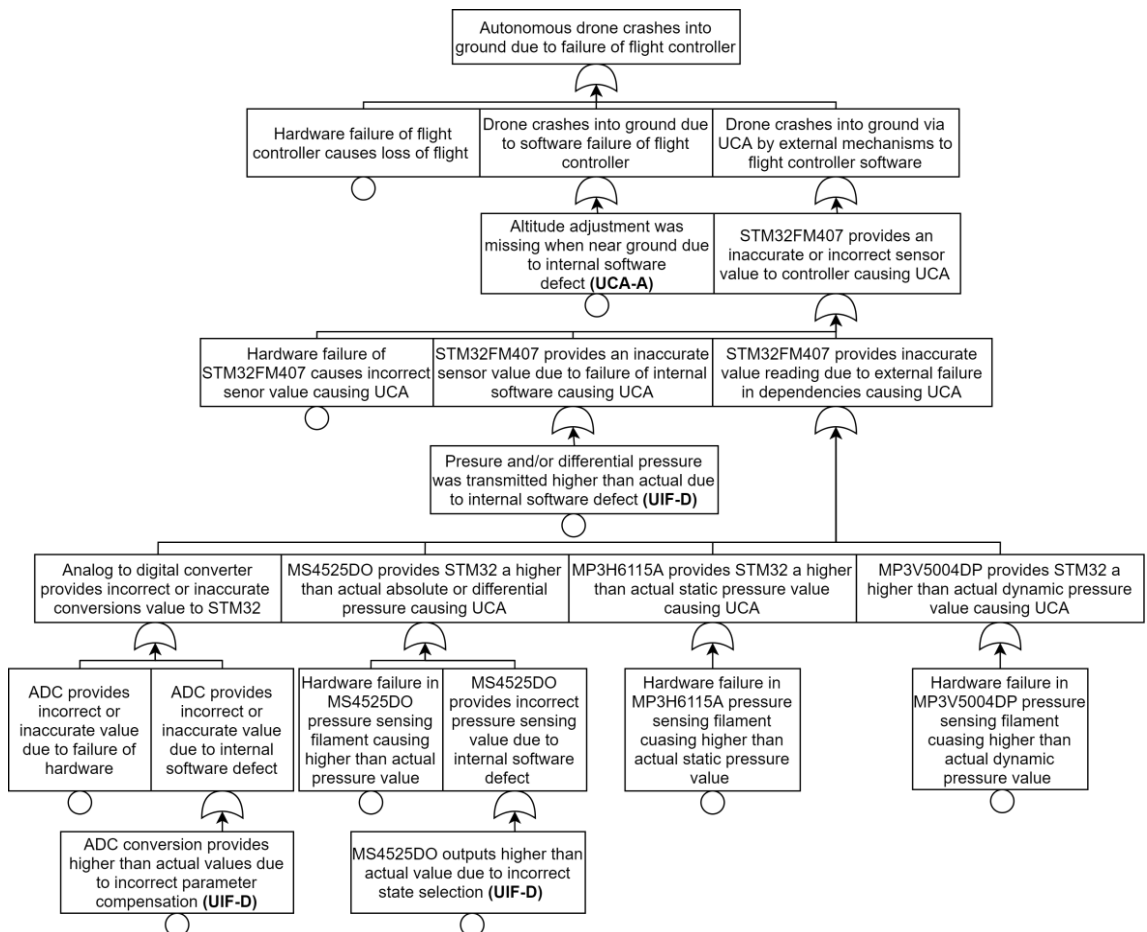

Figure 5. Full fault tree for VCU smart sensor device with failure mode and mechanism tracing.

## 4.2 HSI Integrated Fault Tree Analysis

In this case study, only the redundancies within the QIAS-P HSI were considered. Previous work by the author on this particular HSI system was conducted [15], however is further elaborated upon in this work. From documentation, the system consists of programmable logic controllers (PLCs). PLCs are general DI&C systems that can execute any program instructions. In this work, it is assumed that one PLC implements one function. The potential top event relevant to reactor trip safety is, 'Reactor fails to trip when needed causing core damage.' From [6], the following major functions were identified as observed in Table 3: the variables for the heated junction thermocouple (HJTC), reactor vessel level (RVL), reactor coolant saturation margin (RCSM), core exit temperature (CET), and margin until inadequate core cooling (ICC). In essence, the system consists of value calculators that modify the input variables from the raw sensory information, and an alarm system that reads the plant state and alerts the operator to abnormal conditions. Note that all are considered intermediate processors as they do not have control authority. The abridged control structure diagram can be seen in Figure 6.

As mentioned previously, UIFs must be traceable to UCAs to have a meaningful impact to the system. In this case, the only controller in the entire system is the human operator. A possible UCA that can lead to the top event is, 'Operator does not engage undervoltage trip breakers (A1, A2, B1, or B2) under abnormal operational conditions (UCA-A) due to interface confusion.' Note, for humans, the software failure branch is replaced with 'procedural failure.' The cause of this event could exist within the QIAS-P subsystems. For instance, the 'RCSM alarm does not alert the operator when the available coolant saturation margin decreases below the setpoint (UIF-A).' This could be an internal failure, where the setpoint was incorrectly set, or an external issue where one of the dependent sensors or intermediate processors has failed. In the abridged control diagram seen in Figure 6, the RCSM Alarm depends on the RCSM calculator, the HJTC ADC, the CET ADC, as well as the corresponding sensors.

**Table 3. Major digital components identified within one division of the HSI system.**

| Component ID | Function | Digital Output |
|---|---|---|
| ICC Calculator | Uses HJTC temperature, RVL, RCSM, and CET to determine margin until ICC. | Margin until ICC |
| HJTC temp. Calculator | Takes differential HJTC data and calculates to temperature. | Core temperature |
| RVL Calculator | Takes differential HJTC data, hot and cold leg, pressure, and vessel head to determine coolant level. | Coolant level |
| RCSM Calculator | Takes differential HJTC and absolute CET digital data and converts to coolant saturation margin. | Coolant saturation margin |
| CET Calculator | Takes raw CET digital data and calculates coolant core exit temperature. | Coolant exit temperature |
| ICC Alarm | Compares the available ICC margin with setpoint, triggers alarm for operator if exceeding. | Boolean/alarm to operator |
| HJTC Alarm | Compares the true core temperature with setpoint, triggers alarm for operator if exceeding. | Boolean/alarm to operator |
| RVL Alarm | Compares the available coolant level with setpoint, triggers alarm for operator if exceeding. | Boolean/alarm to operator |
| RCSM Alarm | Compares the available coolant saturation margin with setpoint, triggers alarm for operator if exceeding. | Boolean/alarm to operator |
| CET Alarm | Compares the true core exit temperate with setpoint, triggers alarm for operator if exceeding. | Boolean/alarm to operator |
| HJTC ADC | Converts the analog HJTC voltage readings from the core sensors to digital logic. | Raw core temperature |
| CET ADC | Converts the analog CET voltage readings from the upper plenum to digital logic. | Raw coolant exit temperature |

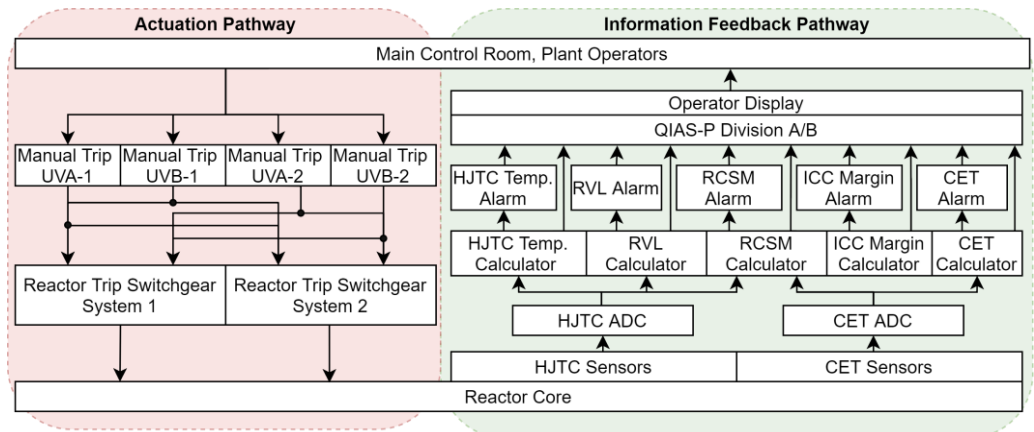

Figure 6. Control diagram of one redundant division of the HSI with manual reactor trip actuation pathway. Left side are control and actuation signals, right side are feedback signals.

The abridged fault tree can be seen in Figure 7. In the full fault tree, a total of 43 minimal software cut sets were discovered. Three representative sets are listed in Table 4 as examples. In cut set #1, a latent internal software failure in the HJTC value calculator results in a nominal (but lower) than real state reading (UIF-D). This can cause both an inter- and intra-division level CCF. As both divisions A and B are assumed to be redundant and not diverse, a software defect would impact HJTC value calculators in both divisions. Furthermore, the output from the HJTC calculator is also used in the HJTC temperature alarm and margin for ICC calculator and alarm, potentially causing an intra-division CCF. The multiple dependent subsystems would also reflect safer than true state. The top event is thus triggered as the operator is unaware the software has failed, and the reactor is hotter than anticipated. A similar scenario can occur for cut set #2, where the lack of diversity among divisions makes the alarm system susceptible to CCFs via software defects—whether by design or by implementation. In cut set #3, a software failure of the ADC would impact the most components in the system. Of the twelve components, nine would experience an intra-division CCF, specifically a failure due to dependency on a singular source. This result was not unexpected.

The primary causes of CCF in the system can be traced back to the lack of diversity by the two QIAS-P redundant divisions. From the documentation, there is sufficient validity to this argument as it is neither explicitly nor implicitly implied that design diversity is considered [6]. The justification for the original designers was that the IPS and DIS would act as diverse backups to the QIAS-P should a software CCF be speculated. However, the IPS is not listed as a safety critical system, monitors different sensor parameters to the QIAS-P system, and is not originally intended for accident scenarios. Furthermore, the DIS, while available in speculated CCF, has significantly reduced capabilities than the QIAS-P. Relevant key safety parameters, such as RCSM, are not determined by the DIS. This suggests, that as the first primary source of safety information to the operator, diversity among QIAS-P divisions should be considered or a secondary monitoring system should be considered safety critical (i.e., the IPS). These are examples of recommendations that can be made to the designers to prevent issues identified in the system (not applicable to the real QIAS-P). Risk assessment of identified CCF events should also be conducted to make adequate risk-informed recommendations.

Table 4. Minimal software cut sets.

| # | Cut Set / Basic Event Description |
|---|---|
| 1 | Division A&B HJTC calculators provide lower than real core temperature measurements under abnormal conditions (UIF-D). |
| 2 | Division A&B CET alarm does not trigger when coolant exit temperature exceeds acceptable margin (UIF-A). |
| 3 | Division A&B HJTC ADC provide lower than real digital core temperature conversion from analog signal (UIF-D). |

## 5. CONCLUSION

In this work, a new method for formulating fault trees to include information pathways is proposed. Unlike conventional methods, the addition of intermediate processors and UIFs allow analysts flexibility when it comes to assessing failure mechanisms in the information pathway. No existing methods were discovered that trace failures in the information system, which is why the authors believe these concepts will be useful to safety analysts. The method was demonstrated on two completely different systems. The first was a smart digital sensor with a lightweight operating system and peripherals. The second was a representation of the APR-1400 QIAS-P, a highly complex HSI with no control authority. Integrated fault trees were developed for both DI&C systems to assess the potential hardware and software failure modes and mechanisms for dependent devices. Future work includes the quantification of the identified UIFs and methods to gather evidence for software reliability.

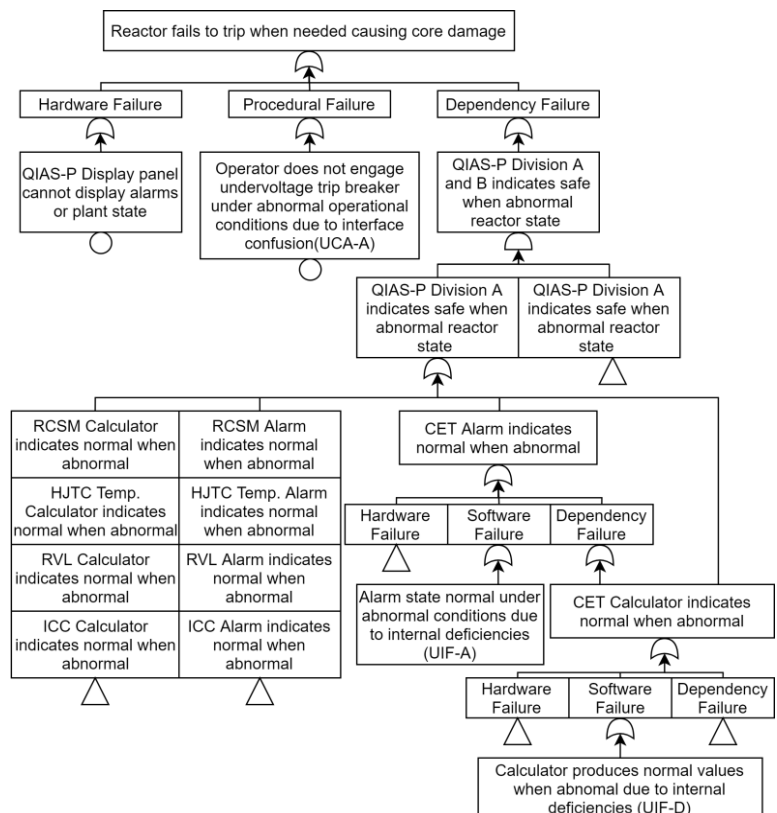

Figure 7. Abridged integrated fault tree of one redundant division of the HSI with UCA/UIF failures.

**Acknowledgments**

The authors would like to thank Gordan E. Holt at Idaho National Laboratory for technical editing and formatting of this paper as well as Dr. Sai Zhang for technical review. This submitted manuscript was authored by a contractor of the U.S. Government under DOE Contract No. DE-AC07-05ID14517. Accordingly, the U.S. Government retains and the publisher, by accepting the article for publication, acknowledges that the U.S. Government retains a nonexclusive, paid-up, irrevocable, worldwide license to publish or reproduce the published form of this manuscript, or allow others to do so, for U.S. Government purposes. This information was prepared as an account of work sponsored by an agency of the U.S. Government. Neither the U.S. Government nor any agency thereof, nor any of their employees, makes any warranty, express or implied, or assumes any legal liability or responsibility for the accuracy, completeness, or usefulness of any information, apparatus, product, or process disclosed, or represents that its use would not infringe privately owned rights. References herein to any specific commercial product, process, or service by trade name, trademark, manufacturer, or otherwise, does not necessarily constitute or imply its endorsement, recommendation, or favoring by the U.S.

**References**

[1] Nuclear Regulatory Commission, "Guidance for Evaluation of Defense in Depth and Diversity to Address Common-Cause Failure due to Latent Design Defects in Digital Safety Systems, BTP 7-19," Nuclear Regulatory Commission, Washington, 2021.

[2] N. G. Leveson and J. P. Thomas, "STPA Handbook," MIT Partnership for Systems Approaches to Safety and Security, Cambridge, 2018.

[3] Electric Power Research Institute, "Hazard Analysis Methods for Digital Instrumentation and Control Systems - Revision 1," Electric Power Research Institute, Washington, 2021.

[4] H. Bao, T. Shorthill and H. Zhang, "Hazard analysis for identifying common cause failures of digital safety systems using a redundancy-guided systems theoretic approach," *Annals of Nuclear Energy,* vol. 148, no. 1, p. 107686, 2020.

[5] T. Shorthill, H. Bao, H. Zhang and H. Ban, "A Redundancy-Guided Approach for the Hazard Analysis of Digital Instrumentation and Control Systems in Advanced Nuclear Power Plants," *Nuclear Technology,* vol. 208, no. 5, pp. 892-911, 2022.

[6] Korea Electric Power Corporation, Korea Hydro & Nuclear Power Co., LTD, "Chapter 7: Instrumentation and Controls. Rev 3," in *APR1400 Design Control Document Tier 2*, Washington, Nuclear Regulatory Commission, 2018.

[7] H. Bao, H. Zhang and K. Thomas, "An Integrated Risk Assessment Process for Digital Instrumentation and Control Upgrades of Nuclear Power Plants," Idaho National Laboratory, Idaho Falls, 2019.

[8] H. Bao, T. Shorthill and H. Zhang, "Redundancy-guided System-theoretic Hazard and Reliability Analysis of Safety-related Digital Instrumentation and Control Systems in Nuclear Power Plant," Idaho National Laboratory, Idaho, 2020.

[9] H. Bao, T. Shorthill, E. Chen and H. Zhang, "Quantitative Risk Analysis of High Safety-significant Safety-related Digital Instrumentation and Control Systems in Nuclear Power Plants using IRADIC Technology," Idaho National Laboratory, Idaho Falls, 2021.

[10] D. J. Lawson, "Failure Mode, Effect and Criticality Analysis," in *Electronic Systems Effectivess and Life Cycle Costing*, Berlin, Springer, 1983, pp. 55-74.

[11] A. D. Williams and A. J. Clark, "Using Systems Theoretic Perspectives for Risk-Informed Cyber Hazard Analysis in Nuclear Power Facilities," in *29th Annual INCOSE International Symposium*, Orlando, 2019.

[12] The National Aeronautics and Space Administration, "Fault Tree Handbook with Aerospace Applications," The National Aeronautics and Space Administration Office of Safety and Mission Assurance, Washington, 2002.

[13] H. Bao, H. Zhang, T. Shorthill and S. Lawrence, "Quantitative Evaluation of Common Cause Failures in High Safety-significant Safety-related Digital Instrumentation and Control Systems in Nuclear Power Plants," 7 April 2022. [Online]. Available: https://arxiv.org/abs/2204.03717. [Accessed 2022].

[14] C. Elks, C. Deloglos, A. Jayakumar, A. Tantawy, R. Hite and S. Guatham, "Specification of a Bounded Exhaustive Testing Study for a Software-based Embedded Digital Device," Idaho National Laboratory, Idaho, 2018.

[15] G. D. Sirio, "ChibiOS/RT The Ultimate Guide," ChibiOS EmbeddedWare, 2020. [Online]. Available: https://www.chibios.org/dokuwiki/doku.php?id=chibios:documentation:books:rt:start. [Accessed 2022].

[16] E. Chen, H. Bao, H. Zhang, T. Shorthill and N. Dinh, "Systems-theoretic hazard analysis of digital human-system interface relevant to reactor trip," in *12th Nuclear Plant Instrumentation, Control and Human-Machine Interface Technologies*, 2021.